\documentclass[12pt]{article}
\usepackage{fullpage}
\usepackage{epsfig}

\def\be{\begin{equation}}
\def\ee{\end{equation}}
\def\u{\underbar}

\begin{document}

\begin{center}
\Large{\bf Bhabha's Contributions to Elementary Particle Physics and
Cosmic Rays Research} \\
\bigskip\bigskip
\large{Virendra Singh} \\
\bigskip
Tata Institute of Fundamental Research \\
Homi Bhabha Road, Mumbai 400 005, India 
\end{center}
\bigskip

\vfill

\noindent $^\star$ email: vsingh1937@gmail.com

\newpage

\noindent {\large\bf 1. Introduction}
\bigskip

Homi Jehangir Bhabha (1909-66) started his career as a theoretical
physicist at Cambridge in the nineteen-thirties and distinguished
himself in the emerging areas of elementary particles, or high energy
physics, and cosmic rays.  He worked actively in these areas for a
period of about two decades (1933-1954) and made many significant
contributions.  His name is commemorated in these fields through the
electron-positron scattering process being called Bhabha
Scattering$^1$.  Bhabha-Heilter cascade theory of cosmic ray
showers$^2$ was a signal contribution to the cosmic ray studies.  His
name is also associated to Bhabha-Corben theory of relativistic
spinning classical particles$^3$.  The Bhabha equations$^4$
investigated possible relativistically invariant linear wave equations
for elementary particles.  Bhabha was also responsible for the
nomenclature `meson' for the bosonic elementary particles of
`intermediate' mass, along with N. Kemmer and M.H.L. Pryce$^5$.  The
terminology `orthochronous' for those Lorentz transformations, which
do not change the sign of time-coordinate, also originated with him in
his work on Bhabha equations$^6$.  Bhabha also initiated experimental
work on cosmic rays, when he was at the Indian Institute of Science,
Bangalore and continued it further at Tata Institute of Fundamental
Research, Bombay (now Mumbai).
\bigskip

We propose to discuss here his various scientific contributions.
\bigskip

\noindent {\large\bf 2. Cambridge Period}
\bigskip

Bhabha went to Cambridge in 1927 and joined Gonville and Caius College
as a student of mechanical engineering.  Around the time the excitment
was rather high there in the theory of elementary particles and
radiation especially in quantum electrodynamics.
\bigskip

Dirac, at Cambridge, had proposed his quantum theory of emission and
absorption of radiation in 1927.  This was followed by his Dirac
equation.  This described relativistically the electron in the same
sense as Maxwell equation describe the radiation.  Dirac equation was
seen to incorporate the phenomenon of electron spin and magnetic
moment quite naturally and was shown to lead to experimentally
correct Sommerfeld formula for the energy levels of the Hydrogen
atom. The main new feature of the Dirac equation was the existence of
negative energy states for electrons and this was rather puzzling as
electrons in these states were predicted to move in a direction
opposite to that of the external applied electromagnetic force.  To
get out of this quandary, Dirac proposed, in 1930, that all the
negative energy states are filled with one electron each in accordance
with Pauli principle.  An unoccupied negative energy state, i.e. `a
hole', would behave as a charged particle with an electric charge
opposite to that of an electron.  This `hole' theory of Dirac was the
first introduction of the concept of `antiparticles' in physics.  These
particles, now called positrons, were observed experimentally by
C.D. Anderson in 1932 and have the same mass as electron.
\bigskip

Bhabha was swept was all this intellectual excitement in quantum
electrodynamaics.  He wrote to his father asking his permission to
change to theoretical physics.  As he wrote, `I seriously say to you
that business or a job as an engineer is not the thing for me.  It is
totally foreign to my nature and radically opposed to my temperament
and opinions.  Physics is my line, I know I shall do great things
here'.  His father was agreeable to let him pursue theoretical
phyiscs, and get a mathematics tripos, provided Bhabha devoted himself
first to his mechanical tripos, and obtained a first class.  Bhabha
did that in June 1930 was free to devote himself to his interest in
theoretical physics thereafter.  His initial interests were mainly in
the positron theory and cosmic rays physics.  We shall describe some
of his main work in these areas before going on to describe his later
work in meson theory.
\bigskip

\noindent {\large 2.1 \u{Positron theory: Bhabha Scattering}}
\bigskip

Bhabha investigated positron interactions in a number of his papers
using Dirac's hole theory.  The first of these dealt with
``Annihilation of Fast Positrons by Electrons in the K-shell'' which
was received by Royal Society on May 4, 1934.$^7$  The work on this
paper was begun at Cambridge with H.R. Hulme, and finished by Bhabha
at Institute of Physics in Rome where he was visiting Enrico Fermi on
his Rouse Ball Travelling studentship in Mathematics which he held for
1932-1934.  The next electrodynamic process he considered was the
creation of electron-positron pairs by fast charged particles$^8$.  He
was holding an Isaac Newton studentship (1934-1936) at this time.
\bigskip

The crowning achievement of this group of papers on positron
interactions was Bhabha's investigation of electron-positron
scattering, a process now known as ``Bhabha Scattering''.  The paper
is titled `The Scattering of Positrons by Electrons with Exchange on
Dirac's Theory of Positron'$^1$ and was received by the Royal Society
on Oct. 20, 1935.
\bigskip

Mott had earlier considered exchange effects in non-relativistic
electron-electron scattering.  He had found that exchange effects are
considerable except that the effect vanishes when the `two electrons
have antiparallel spins'$^9$.  M\"oller had generalised these results
to the relativistic electron-electron scattering, now called M\"oller
Scattering$^{10}$, and found that the exchange effects are
non-vanishing even when electron spins are antiparallel except in the
non-relativistic limit.
\bigskip

Now if a positron is regarded as an independent particle, which also
obeys Dirac equation, then the positron-electron scattering should
show no exchange effects.  If on the other hand, positron is regarded
as an electron in an unoccupied negative energy state then we should
expect exchange effects.  These two hypotheses would lead to different
results.  To quote Bhabha, ``The difference would be due to the effect
of exchange between the electron we observe initially and the virtual
electrons in states of negative energy''$^{11}$.  Except in the
non-relativistic limit, the effect of the exchange was found
considerable.  It was not completely clear at this time whether such
an exchange effect is simply not an incorrect prediction of the Dirac
theory as the exchange was between an observable electron of positive
energy and another virtual one of negative energy.
\bigskip

Bhabha however pointed out that another way of looking at this extra
exchange contribution was to regard it as due to annihilation of an
electron-positron pair, followed by simultaneous creation of a new
electron-positron pair and that such terms should be present in the
scattering of any two particles which can annihilate each other and be
created in pairs.  He noted that such terms were present in a recent
theory of Pauli and Weisskopf$^{12}$ having particles which obey not
the exclusion principle but rather obey Einstein-Bose statistics.
\bigskip

A consequence of this calculation was an expected considerable
increase in the number of fast secondaries for positrons of high
energies.  Bhabha's theory was beautifully confirmed by experiments.
Our faith in the correctness of Bhabha scattering formulas is such
that they are now routinely used to calibrate the beams at large
accelerators using positron or other antiparticle beams.
\bigskip

It should be noted that the discovery of substitution law or crossing
symmetry property of local quantum field theory allows us to get the
matrix elements for Bhabha Scattering from those of the M\"oller
scattering and vice versa$^{13}$.  Of course the discovery of the
substitution law itself owes much to these earlier calculations of the
related processes.
\bigskip

\noindent {\large 2.2 \u{Cosmic Ray}}
\bigskip

In 1936, Bhabha was awarded a Senior Studentship of the Exhibition of
1851.  This was also the year in which he and W. Heitler, from
Bristol, met together.  They had common interest in high energies.
Indeed the very first publication of Bhabha in 1933, written from
Zurich where he was visiting W. Pauli from Cambridge, was `On the
Absorption of the Cosmic Rays$^{14}$ and he had discussed the role of
electron showers in it.  Since that time he had devoted himself to
positron processes in quantum electrodynamics.  W. Heitler had also
been working on similar subjects.  Bethe and Heitler had worked on
bremstrahlung radiation from charged particles and on high energy
photon induced pair production.  Heitler had published the first
edition of his book ``Quantum Theory of Radiation'' in 1936.  The
result of this collaboration between Bhabha and Heitler was their
celebrated cascade theory of electron showers$^2$.
\bigskip

\noindent {\large 2.2.1 \u{Bhabha-Heitler Theory}}
\bigskip

The Bethe-Heitler theory predicted large crosssections for energy loss
of electrons or positrons passing through the field of a nucleus by
bremstrahlung in a {\it single} encounter. These hard quanta also have
a large probability for materialising as electron-positron pairs.  The
theoretical ``range'' of an electron of $10^{12}$ eV was found to be
only about 2 km of air, 2 m of water or 4 cm of lead.  It, however,
seemed to disagree with the observation of fast electrons at sea level
which have traversed 8 km of atmosphere.  The common belief was that
these observations signify a breakdown of Quantum Electrodynamics.
Bhabha and Heitler however went on to show, through their cascade
theory, that quantum electrodynamics was quite consistent with the
observed phenomenon.
\bigskip

What happens is as follows: A fast electron does loose all its energy
after a short distance of travelling through matter just as predicted
by Bethe-Heitler formula. This energy however reappears in the form of
radiation quanta which, for large initial electron velocities, have a
large probability of moving in the direction of the original fast
electron.  There is reasonably high probability that these radiation
quanta are hard.  Again, in accordance with quantum electrodynamics,
these radiation quanta materialise, after traversing a short distance
through matter, into electron-positron pairs.  Again there is a
reasonably high probability that the resulting pair consists of a fast
electron and a fast positron moving in the direction of the
disappearing hard quanta.  This process of conversion into photons and
reconversion into electron-positron pairs can take place many times.
The total effect is thus as if the original electron was losing its
energy much more slowly than implied by the theoretical ``range''
mentioned earlier.
\bigskip

These ideas of Bhabha and Heitler provided a natural explanation of
the cosmic ray showers. The calculated curves for the expected number
of electrons $n$ $(E,h;E_0)$, above a certain energy $E$ and at
distance $h$ below the top of the atmosphere when an electron with
energy $E_0$ is incident at the top of the atmosphere, agreed well
with the observed curves by Rossi$^{15}$ and the ionisation curves of
Regener$^{16}$. 
\bigskip

The idea that cosmic ray showers could be explained this way was `in
the air'.  L. Nordheim, in a conversation with Heitler in 1934, and
Carmichael in a conversation with Bhabha had already considered the
possibility but ``owing to the ill-founded suspicion in which the
theory was then held, it did not seem worthwhile carrying out any
calculations''$^{17}$. The cascade theory was also given at about the
same time by J.F. Carlson and J.R. Oppenheimer$^{18}$.
\bigskip

\noindent {\large 2.2.2 \u{Penetrating Component of Cosmic Radiation}}
\bigskip

It was very suggestive from the work on absorption of cosmic rays in
lead by Rossi and others that cosmic rays consist of two components: a
soft component and a penetrating component$^{19}$.  The intensity of
the soft component is reduced by about 30 percent in passage through
about 10 cm of lead.  There are, however, also single charged
particles, belonging to the penetrating component, which can traverse
one meter of lead.  The properties of the soft component could be
completely accounted for, by Bhabha-Heitler theory, if it consisted of
electrons.  The penetrating component was, however, still an enigma.
It was not possible for quantum electrodynamics to account for the
penetrating component if these particles were electrons.  There was
therefore a breakdown of quantum electrodynamics for higher energies
and/or the penetrating components was not composed of electrons.
\bigskip

Bhabha, in a remarkable paper submitted to the Royal Society on
Oct. 4, 1937, carried out a powerful analysis of both the theoretical
and the observational situation about the penetrating component of
cosmic radiation$^{20}$.
\bigskip

He first showed that a ``breakdown'' of quantum electrodynamics for
the energy loss of electrons was not a viable alternative for the
observed phenomenology of cosmic radiation.  He showed that if the
breakdown energy is around 10 GeV or higher for electrons then no
latitude effect would be there at sea level.  The shape of the
transition curve for large showers also led to a similar conclusion.
\bigskip

It therefore seemed that a viable alternative is to regard the
penetrating component of cosmic radiation as consisting of particles
other than electrons.  Could they be protons?  Initially Bhabha and
Heitler had inclined to this view while Blackett was of the view that
they are electrons.  It, however, gradually became clear that these
particles could not be protons either in view of Blackett's arguments
against the proton hypothesis$^{21}$.
\bigskip

Bhabha therefore investigated in this paper the hypothesis that
``There are in the penetrating component of Cosmic radiation new
particles of electronic charge of both signs, and mass intermediate
between those of the electron and proton''$^{22}$.  He called them heavy
electrons.  The illustrative mass values considered were 10 and 100
times the electron mass.  We quote ``A comparison with the
measurements of Auger and others therefore allows one to conclude that
if the hypothesis of new particles is right, {\it the majority of the
penetrating particles must have masses nearer to a hundred times the
electron mass rather than ten times the same}''$^{23}$.  Thus Bhabha
had, in a brilliant piece of cosmic ray phenomenology, essentially
predicted the existence of a new particle.  Similar conclusions had
also been reached by Neddermyer and Anderson, and by Street and
Stevenson$^{24}$.  These particles are now called muons and the first
clear case of their track was observed in Cosmic rays in 1938 by
Anderson and Neddermyer$^{25}$.  Their mass is about two hundred
electron masses.
\bigskip

\noindent {\large 2.3 \u{Meson Theory}}
\bigskip

Bhabha, when he predicted the ``meson'' in October 1937, was not aware
of Yukawa's meson theory$^{26}$, which was brought to his attention by
Heitler in a discussion about cosmic radiation presumably some time
between October 4, 1937 and Dec. 13, 1937 when he sent his first note
on meson theory for publication$^{27}$.  It was natural for Bhabha to
identify Yukawa's mesons with those indicated by cosmic ray
phenomenology. 
\bigskip

\noindent {\large 2.3.1 \u{A Test of Relativistic Time-dilation}}
\bigskip

It was first pointed out by Bhabha, in his note to Nature$^{27}$ that
positive (negative) mesons should sponteneously decay into a positron
(electron).  In this note he also pointed out ``This distintegration
being spontaneous, the $U$-particle may be described as a `clock', and
hence it follows merely from considerations of relativity that the
time of distintegration is longer when the particle is in
motion''. The $U$-particles referred to the mesons.  Thus the lifetime
of a particle $T$, moving with a velocity $v$, should be given by 
\[
T = T_0/\sqrt|1 - (v/c)^2|
\]
where $T_0$ is the lifetime of the particle at rest and $c$ is the
velocity of light.  This was nicely confirmed by experiments and
constitutes one of the most beautiful tests of the special theory of
relativity. 
\bigskip

\noindent {\large 2.3.2 \u{Vector-Meson Theory}}
\bigskip

The identification of Yukawa's mesons with those required by cosmic
ray phenomenology allows one to couple the problem of nuclear forces
with that of the cosmic rays.  It may be noted that the particles
required by cosmic ray phenomenology are now identified with muons,
while the Yukawa mesons are identified with the pions which were only
discovered in 1948 by C.F. Powell.  However this confusion of the
Yukawa mesons with muons was very fruitful as it gave a great impetus
to the development of the meson theory which had not received much
attention till that time.  The earliest major workers on Meson Theory,
apart from those in Yukawa's group i.e. S. Sakata and Taketani, were
Bhabha at Cambridge, N. Kemmer in London, H. Fr\"ohlich and W. Heitler
at Bristol.
\bigskip

The original Yukawa theory had considered the mesons to be scalar
particles, i.e. having no spin and with positive parity.  This
assignment does not give rise to a satisfactory nuclear force between
nucleons i.e. protons and neutrons.  Mesons, further, have to have
integral spin and must obey Bose-Einstein statistics.  Bhabha
therefore considered the generalisation that the mesons are vector
particles i.e. having spin one and odd parity.  He used Proca's wave
equation to describe the meson field.  The coupled nucleon-meson field
system was then quantised and nucleon-nucleon interaction calculated
using second order perturbation theory.  Bhabha was gratified that
``The interaction is therefore just of the required form consisting of
Heisenberg and Majorana forces of the right sign so as to allow one to
make the triplet state of the deuteron the lowest stable state.  We
would emphasize the fact that since only the squares of $g_1$ and
$g_2$ enter into this expression, the sign of the Majorana force is
beyond our control, and it is to be looked upon as a strong argument
in favour of this theory that it allows only that sign of the force
which actually occurs in nature''$^{28}$.  Bhabha also calculated
meson-nucleon scattering crosssections in the lowest order.  There
were at that time no data for this process.  Related investigations
were also carried out by other workers.
\bigskip

\noindent {\large 2.3.3 \u{Classical Meson Theory}}
\bigskip

The vector nature of the meson fields gave rise to large probabilities
at high energies for multiple processes.  Also the theory had more
severe divergence problems compared to quantum electrodynamics.  There
were apprehensions that these imply the breakdown of either quantum
mechanics or meson theory even for lower energies comparable to meson
mass.  All of these were essentially based on second order
perturbative calculations in meson-nucleon coupling constants.  It is
to be remembered that these coupling constants are rather large as the
meson-nucleon interaction is strong.  As such these calculations are
not necessarily reliable.
\bigskip

In order to escape the limitations of the perturbation theory in
coupling constant, in dealing with quantum meson-field theory, Bhabha
decided to investigate the classical meson field theory in interaction
with a fermion$^{29}$.  The meson field was described, as earlier, by
the Proca wave equation, except that now the meson field components
were taken as commuting variables.  The fermion was taken as a point
classical particle having spin and moving along classical world-lines.
This entailed a generalisation of the method of Dirac used in treating
the behaviour of a classical point electron in the field of
electromagnetic radiation$^{30}$.
\bigskip

The meson-fermion scattering crosssections were calculated.  These are
analogous to Thomson scattering for zero mass photons and were found
to smoothly go to that limit as meson mass went to zero.  Indeed it
was found that at high energies, i.e. energies much larger than meson
mass, the behaviour was essentially the same as that in Dirac's case.
Before the advent of the Chew-Low theory, these were among the best
theoretical attempts to deal with the meson-nucleon scattering
problem.
\bigskip

\noindent {\large\bf 3. Bangalore Period}
\bigskip

Bhabha was in India, in 1939, on a holiday when the second world war
broke out.  The subsequent war conditions made it impossible for him
to return to England.  His earliest communications to the Proceedings
of the Indian Academy of Sciences are from this period and were
received by them in October 1939.  The second of these two papers
deals with the classical theory of the electron and is still bylined
as Gonville and Caius College.  In early 1940 Bhabha joined Indian
Institute of Science, Bangalore as special Reader incharge of a Cosmic
Ray Unit.
\bigskip

\noindent {\large 3.1 \u{Classical Relativistic Spinning Point
Particle Theory:} 

~\u{Bhabha-Corben equations}}  
\bigskip

Bhabha was originally drawn to the classical theory of relativistic
point particles as a way to take into account the reaction of the
emitted or scattered radiation, whether of electromagnetic quanta or
of mesons i.e. radiation reaction, on the motion of fermions.  The
quantum treatment of the interaction of point particles with fields,
which depended on the perturbation theory in the coupling constant,
was not very satisfactory.  It was even more so when the explicit spin
dependant interaction, e.g. Pauli anomalous moment term for
interaction with electromagnetic field or similar for vector meson
fields, were taken into account.  The interactions tend to increase
with energy.  It was pointed out by Bhabha that these effects are due
to neglect of the radiation reaction and that the quantum treatment
can be trusted only in the region of energy where this neglect of
radiation reaction is justified.  Faced with this situation it was
natural to go back to the classical limit where it is, in principle,
possible to take radiation reaction into account either exactly or
with controlled approximations.  For spinless charged particles it was
possible to do this by using Dirac's work on the point electron
theory$^{30}$.  The electron theories had been in existence since
Lorentz's work in 1892 but Dirac's work of 1938 was the first
logically correct relativistic classical point electron theory.
\bigskip

In Dirac's work the spin of the electron was not taken into account.
It was therefore necessary to generalise Dirac's work to the case of
spinning point particles.  Bhabha had already begun some work on this
problem with H.C. Corben before leaving Cambridge and it was continued
at Bangalore.  A preliminary note on this work was sent to Nature on
March 17, 1940 and was his first research note from Indian Institute
of Science, Bangalore.  The definitive paper on this work, giving
Bhabha-Corben equations, was titled ``General Classical Theory of
Spinning Particles in a Maxwell field''$^3$.  The case of a meson
field was treated in a sequal which followed immediately after this
paper$^3$. 
\bigskip

The point electron theories previous to the work of Dirac, and related
work of Pryce, had approached the problem of a point electron as the
limit of a finite-size electron.  This procedure is not very
satisfactory in view of the conflict between the concept of a
rigid-body and relativity.  In the procedure of Dirac and Pryce one
assumes the validity of field equations right upto the point particle
world-line but one modifies the definition of the field energy in the
presence of singularities.  This is also the procedure adopted by
Bhabha and Corben.
\bigskip

The effect of radiation reactions was, as expected by Bhabha, to
reduce the scattering crosssection.  Indeed for large energies it was
found to decrease inversely as the square of the incident photon
frequency for the compton-process.
\bigskip

Bhabha was elected a Fellow of the Royal Society, London in 1941.
\bigskip

\noindent {\large 3.2 \u{Meson Theory and Nucleon Isobars}}
\bigskip

Before the discovery of the pion in 1947 the muon was confused with
Yukawa's meson. The meson theorists had the unenviable task of
explaining as to why the meson-nucleon scattering is weak -- since
muons do not scatter much on the nucleons -- and yet at the same time
the nuclear forces arising from the exchange of the same mesons is
strong. 
\bigskip

It was noted by Bhabha that the scattering of the longitudinally
polarised {\it neutral} vector mesons on nucleons shows a decrease as
$E^{-2}$, $E$ being the meson energy, for large energies in contrst to
the scattering of longitudinally polarised {\it charged} mesons whose
scattering on nucleons increases as $E^2$.  This difference was traced
by Bhabha to what we would refer in modern terminology as being due to
the cancellation between direct channel nucleon pole and the crossed
channel nucleon pole in case of neutral mesons and a lack of such a
cancellation for charged mesons.  For charged mesons we do not have
both direct and crossed channel exchanges possible if only Nucleons of
charge +1 and 0 exist.  In order to have the cancellation mechanism
available, Bhabha therefore suggested that the nucleon may exist in
charge states +2 and -1 also.  The contribution from these charged
states was needed to provide the cancellation. In general nucleon
isobars may have any charge and neutron and proton are only the
lightest ones occuring in nature$^{31}$.  This was the first
suggestion of the existence of the nucleon isobars.
\bigskip

Bhabha communicated the idea to Heitler and he also pursued it.  This
mechanism, referred to as Bhabha-Heitler mechanism, for reducing the
mesons crosssections, was one of the major reasons for a study of the
strong coupling theory of nucleon isobars.  The first nucleon isobar
$N^\star \ (1240 \ {\rm MeV})$ was discovered by Fermi et al in the
pion-nucleon scattering experiments in 1952.
\bigskip

\noindent {\large 3.3 \u{Cosmic Rays}} 
\bigskip

In his capacity as special reader in charge of cosmic ray unit at
Indian Institute of Science, Bangalore, Bhabha planned to pursue both
theoretical and experimental work in the area of cosmic rays.  The
unit was set up as part of the department of Physics which was headed
by Nobel Laureate Sir C.V. Raman.  Given his pioneering work on cosmic
ray showers with Heilter, and the scope offered by the experimental
work on Cosmic to study high energy interaction, involving particle
production, it was a natural choice.  India also offered a tremendous
geographical advantage of comprising magnetic latitudes ranging
extensively from equator to 25$^\circ$ North within its' confines for
studying high energy cosmic rays.
\bigskip

\noindent {\large 3.3.1 \u{Cascade theory}}
\bigskip

The initial cosmic ray work in the cosmic rays concerned itself with
refinements of the classic Bhabha-Heitler theory.  They had made a
number of simplifying assumptions.  In particular it was assumed that
one can ignore collision loss below a certain critical energy i.e. the
energy at which collision loss is equal to radiation loss.  Further if
the energy is above this critical energy it was assumed that it would
lead to an absorption of the cascade.  In view of the improvement in
the quality of the observational data it had become necessary to
improve the theoretical treatment.
\bigskip

There had been previous attempts, notably by Snyder and Serber, to
give an improved treatment taking collision loss into account$^{32}$,
but there were some doubts about the convergence of their series
solutions.  Bhabha and Chakrabarty gave a solution of this problem in
the form of a rapidly converging series$^{33}$.
\bigskip

In all these treatments the lateral spread of the shower is
neglected.  The collision loss is taken as a constant $\beta$
independent of the energy of the charged particle.  Actually it is not
strictly constant but in the whole relevant energy range of 5 to 150
MeV it increases less than by a factor of 1.5 and it is thus
reasonable to treat it is a constant.  For radiation loss and pair
creation one can use the exact expressions of Bethe and Heitler.  An
exact solution of the Landau-Rumer equations$^{34}$, describing the
longitudinal evolution of the shower, for the number of charged
particles and radiation quanta was given by K.S.K. Iyengar$^{35}$.
His solution was, however, not easily amenable to extracting numerical
results.  Bhabha and Chakrabarty, used the asymptotic form of the
Bethe-Heitler expressions, in their solution.  Using Mellin transform
techniques it was then possible to obtain a series solution which was
rapidly convergent.  It is possible to regard the Bhabha-Chakrabarty
solution as an analytic continuation of the results of Snyder and
Serber. 
\bigskip

\noindent {\large 3.3.2 \u{Experimental cosmic ray work}}
\bigskip

The first problem which Bhabha decided to tackle experimentally in
cosmic rays was to study latitude effect for mesons.  The soft
component of the cosmic rays, consisting of electrons, positrons and
gamma rays, was described quite correctly by the cosmic ray shower
theory.  The hard component of the cosmic rays, consisting mainly of
mesons, was much less understood theoretically and needed experimental
observations.  The hard component would also include high energy
protons. 
\bigskip

In order to devise experimental set up to study the hard component of
the cosmic rays, containing mesons, it was necessary to devise
procedures to discriminate between the soft and the hard components
experimentally.  Bhabha studied this problem in 1943.  He came to the
conclusion, using recent detailed work of Bhabha and
Chakrabarty$^{33}$ on the cosmic ray showers, that the usual method of
separating the two components of cosmic rays by interposing absorbers,
such as lead, of different thickness, and measuring the cosmic ray
absorption, is not sufficiently reliable.
\bigskip

Bhabha devised a new method (Bhabha method) for this purpose.  The
absorper of total thickness $t$ is divided into two of thickness $t_1$ 
and $t - t_1$.  Between the two parts of the absorbers are interposed
a set of counters $C$ and $D$ in anticoincidence.  The whole sandwich
is then placed between the set of counters $A$ on one side and
counters $B$ on the other side.  The set of counters $A,B$ and $C$ (or
$D$) are in coincidence.  Bhabha showed that such an arrangement would
be able to take better advantage of shower multiplication by soft
component to better discriminate the soft and the hard
component$^{36}$. 
\bigskip

The experimental measurements of the vertical intensity of mesons were
carried out in two aeroplane flights at Bangalore with magnetic
lattide of 3.3$^\circ$ N.$^{37}$  The first flight carried two sets of
counter telescopes viz (i) with total thickness of lead absorber equal
to 5.25 cms and divided into two absorber one of 1.25 cm and other of
4.0 cm.  The arrangement here used Bhabha method and (ii) a quadruple
coincidence counter telescope with 3.0 cm of lead absorber.  The
measurements were taken upto 15000 ft.  The second flight carried a
quadrupole coincidence counter with 20 cm of lead and carried out
measurements upto 30,000 ft.  A comparison of the experimental
measurements by Bhabha and his group in these flights with those of
Schein, Jesse and Wollan$^{38}$ carried out at Chicago, magnetic
latitude 52.5$^\circ$ N, showed that for meson intensity there was
``no marked increase even to attitudes corresponding to 275 milli bars
pressure''.  This was quite in contrast to the total cosmic ray
intensity which showed a marked increase of latitude effect upto these
heights. 
\bigskip

Another flight making such measurement later extended the results upto
an attitude of 40,000 ft above Bangalore with similar results.
\bigskip

\noindent {\large\bf 4. Bombay Period}
\bigskip

During the five year period in Bangalore ``he found his mission in
life''$^{39}$ as an institution builder.  As a result of
correspondence with J.R.D. Tata and Sir Sorab Saklatvala, he founded
the Tata Institute of Fundamental Research at Bombay in 1945.  The
institute started functioning in June 1945 with H.J. Bhabha as its
Director and was formally inaugurated on 19 December 1945.  Initially
the work at the Institute was carried out in the areas of mathematics,
theoretical physics and a continuation of the experimental work in
cosmic rays.  He was appointed as the first chairman of the Atomic
Energy Commission when it was founded in 1948 and became Secretary to
the Department of Atomic Energy of the Government of India in 1954.
\bigskip

\noindent {\large 4.1 \u{Bhabha Equations}}
\bigskip

The first research paper from Tata Institute dealt with ``Relativistic
Wave Equations for the Elementary Particles'' and appeared in an issue
of Reviews of Modern Physics to commemorate the Sixtieth birthday of
Prof. Niels Bohr$^4$. 
\bigskip

Dirac, as mentioned earlier, had given his relativistic wave equation
in 1928 which described the behaviour of electrons with spin one half
and had successfully predicted the existence of positrons.  Dirac
equation was a first order equation and its success encouraged similar
attempts to find wave equations describing particles with a spin
having a value other than one half.  Duffin, Kemmer and Petiau had
given similar first order wave equations describing particles with
spin-0 and spin-1. The spin-1 equations were Proca equations. Dirac,  
and Fierz and Pauli had proposed relativistic wave equations for
particles having any integral or half-odd integral spin.  An
unsatisfactory feature of the equations proposed by Dirac, and Fierz
and Pauli for spin greater than one, was the presence of subsidiary
conditions. These subsidiary conditions created difficulties when one
considered these particles in interaction with electromagnetic fields.
The difficulty was connected to the fact that these could not be
derived from a variational principle.  Bhabha therefore proposed to
investigate general first-order relativistic wave equations, without
any subsidiary conditions i.e. equations for wave field $\psi$ of the
form 
\[
(\alpha^k p_k + \chi)\psi = 0
\]
where $p_k = i\partial/\partial x^k$, $\alpha^k$ s are four
matrices $(k = 0,1,2,3)$ and $\chi$ is an arbitrary constant.  Such
equations have come to be known as Bhabha equations.  Sometimes the
nomenclature ``Bhabha equations'' is used to refer to a restricted
subclass of equations where six Lorentz generators, together with four
$\alpha^k$ s, form an $SO(5)$ algebra.  This restricted subclass was
investigated by Bhabha in detail and from now on in this section we
shall mean this subclass when we refer to Bhabha equations.
\bigskip

As Bhabha equations involve an $SO(5)$ algebra they can be completely
classified by using the representation theory of the $SO(5)$ algebra.
Bhabha was among those few physicists of his time who were at home
with group theory.  His essay ``The Theory of the Elementary Physical
Particles and their Interactions'', for which he was awarded Adams
Prize in 1942, had contained a fair amount of the theory of orthogonal
groups.  Bhabha equations were found to contain Dirac equation for
spin one-half and Duffin-Kemmer-Petiau equations for spin-0 and spin-1
as special cases.  These equations were special cases for
Dirac-Fierz-Pauli equations also.  Bhabha equations, for spin greater
than one, were however from Dirac-Pauli equations, and led to multiple
mass states.

\newpage

\noindent {\large 4.2 \u{Cosmic Rays}}
\bigskip

\noindent {\large 4.2.1 \u{Cascade Theory and Stochastic Processes}}
\bigskip

Bhabha and Chakrabarty extended their calculations of the number of
charged particles and quanta, in 1948, to cover the case of showers in
thin layers.  They brought to completion this part of their work on
cascade theory which deals only with mean numbers of shower particles
at various depths.
\bigskip

The study of the fluctuations of the number of shower particles is
also of great importance in the study of cosmic ray showers.  One of
the conceptual difficulties in attacking the problem was that it
involved a system whose state space was continuous and not
discrete. The electrons and photons do have a continuous variation in
energy. Bhabha therefore derived the product density function method
for the continuous parametric systems and applied it to derive the
equations for the cosmic ray cascade theory which determine the mean
numbers (i.e. Landau-Rumer equations) and the mean square deviations
of the numbers$^{40}$.  These equations were solved in a subsequent
paper with Ramakrishnan$^{41}$.
\bigskip

\noindent {\large 4.2.2 \u{Experimental work in Cosmic rays}}
\bigskip

The program of measuring the hard component of the cosmic ray
intensity, at various Indian latitudes and it's variation with
altitudes was continued at Bombay.  Towards this objective Bhabha
organised a High Attitude Studies group whose main program was to
organise Balloon flights for these studies.  The balloon flights were
initially made at from Bangalore and Delhi.  Bhabha reported these
results at a conference at Kyoto-Tokyo in 1953.$^{42}$  Later flights
were also made from many other locations as well.
\bigskip

Bhabha made a preliminary begining for nuclear emulsion work with
cosmic rays by flying in an airplane a Ilford C2 plates loaded with
Boron at an average altitude of 8000 ft for 72 hours in 1948.  An
example of ``meson'' scattering with nuclear excitation was recorded
and published with his student Roy Daniel$^{43}$.  Later Bhabha
induced Bernard Peters, of Rochester University, to take over the
Nuclear emulsion group at Bombay in Dec. 1950.  Peters was well known
for his discovery of heavy nuclei in primary cosmic rays.
\bigskip

A 12$^{\prime\prime}$ diameter cloud chamber, similar in design to one
at Blackett's laboratory at Manchester, which Bhabha had got built at
Bangalore was also moved to Bombay and work on meson scattering
continued.  Prof. Sreekantan took over the further developments
here$^{44}$. 
\bigskip

Bhabha also started thinking about Kolar Gold Fields as a facility for
deep underground experiments on cosmic rays around 1950.
Prof. M.G.K. Menon joined the Cosmic ray group from Prof. Powell's
group in 1956. 

\newpage

\noindent {\large 4.3 \u{Multiple Meson Production}}
\bigskip

This is essentially Bhabha's last piece of theoretical research
work$^{45}$.  His work in Bangalore and Bombay had been decidedly more
mathematical as compared with his work at Cambridge which had been on
the whole, phenomenological.  In his work on multiple meson production
he again displays a phenomenological strain.
\bigskip

In the center of mass system the two colliding nucleons suffer
relativistic contraction in the direction of motion and appear as
colliding discs.  Different mechanisms were invoked by Fermi and
Heisenberg as to how the nucleon energy is converted into mesons.
Bhabha suggested the hypothesis that the strong interactions get
localised both due to relativistic contraction as well as due to the
time of interaction being very small.  On this picture the energy
available for meson production is much less than the total available
c.m. energy.
\bigskip

The real surge of interest in multiple meson production had to wait
till early nineteen seventies when sufficient high energy data became
available.  In Bhabha's picture it is somewhat natural to assume that
not only the strong interaction gets localised but the production of
mesons also gets localised, leading to a damping of transverse
momenta$^{46}$.  Bhabha's model thus can be regarded as a oprecursor
of the parton model.
\bigskip

\noindent {\large\bf 5. Concluding Remarks}
\bigskip

Bhabha's work in theoretical physics was carried out in a wide variety
of styles.  His work, with Heitler, on cascade theory, is of a kind
which would now be described as phenomenology.  Some of his work had
speculative components e.g. his work on penetrating component of
cosmic rays where he suggested the existence of ``muon'' like
particles and his work on meson-nucleon scattering where he suggested
the existence of ``nucleon isobars'' especially those having electric
charge of +2 and -1 in the units of proton charge.  His work on the
theory of relativistic spinning particles in classical physics was
originally motivated by the problem of radiation reaction.  One
cannot, however, help feeling that this motivation was strongly
reinforced by the aesthetic appeal of this investigation.  Here one
has a well defined ``complete'' theory in a world described by
classical physics.  The finer aspects of this theory cannot be tested
in the real world as there are important quantum corrections which
vitiate any testing.  His work on relativistically invariant wave
equations, though motivated by a possible application to ``nucleons'',
could be regarded as almost pure mathematical group theory.  In fact
work on these equations provided the backgorund for the later
important work on the theory of noncompact groups by his collaborator
Harish-Chandra. 
\bigskip

Though Bhabha's main scientific work and achievements were in
theoretical physics.  He was sensitive to the importance of
experimental work as well.  His cosmic ray experiments were not
carried out by using techniques used in similar work elsewhere but
used a noval method devised by himself.
\bigskip

We have restricted ourselves in this paper to the research
contributions of Bhabha in physics.  This hardly exhausts Bhabha's
contribution to science.  He played an important role as a developer
of scientific institutions in India.  He was also an excellent science
administrator with an innovative style of management.  More than any
other person, Bhabha was responsible for introducing and nurturing
modern nuclear science and technology in India.  His role in the
emergence of Bombay school of modern Indian painting has also been
taken note of.  But, in a sense, all these later achievements of
Bhabha have their origins in the excellence he achieved in his
scientific research.
\bigskip

\noindent \underbar{Bibiliographical Notes}:
\bigskip

This paper overlaps in some places with author's earlier related
writings: 
\bigskip

\begin{enumerate}
\item[{(1)}] H.J. Bhabha: His Contributions to Theoretical Physics, in
Homi Jehangir Bhabha: Collected Scientific Papers, Editors:
B.V. Sreekantan, Virendra Singh and B.M. Udgaonkar, p.xxii-xlvii,
TIFR, Mumbai (1985).
\item[{(2)}] H.J. Bhabha in The Scientist in Society, p. 181-193,
Thema, Calcutta (2000).
\end{enumerate}
\bigskip

\noindent {\large\bf References}
\bigskip

\begin{enumerate}
\item[{1.}] H.J. Bhabha, Proc. Roy. Soc. (London) \u{154A}, 195 (1936).
\item[{2.}] H.J. Bhabha and W. Heitler, Proc. Roy. Soc. (London),
\u{159A}, 432 (1937).
\item[{3.}] H.J. Bhabha and H.C. Corben, Proc. Roy. Soc. (London),
\u{178A}, 273 (1941) and \\ H.J. Bhabha, Proc. Roy. Soc. (London),
\u{178A}, 314 (1941).
\item[{4.}] H.J. Bhabha, Rev. Mod. Phys. \u{17}, 200 (1945).
\item[{5.}] N. Kemmer, Suppl. Progr. Theor. Phys., 30th Anniversary of
Meson theory issue, p. 602-608 (1965).
\item[{6.}] H.J. Bhabha, Rev. Mod. Phys. \u{21}, 451 (1949).
\item[{7.}] H.J. Bhabha and H.R. Hulme, Proc. Roy. Soc. (London)
\u{146A}, 723 (1934). 
\item[{8.}] H.J. Bhabha, Proc. Roy. Soc. (London) \u{152A}, 559
(1935), and \\ H.J. Bhabha, Proc. Camb. Phil. Soc. \u{31}, 394 (1934).
\item[{9.}] N.F. Mott, Proc. Roy. Soc. (London) \u{126A}, 259 (1930).
\item[{10.}] C. M\"oller, Ann. Physik \u{14}, 531 (1932).
\item[{11.}] Ref. 1 p. 196.
\item[{12.}] W. Pauli and W. Weisskopf, Helv. Phys. Acta, \u{7}, 709
(1934).     
\item[{13.}] K. Baumann, Acta Physica Austriaca \u{7}, 96 (1953). 
\item[{14.}] H.J. Bhabha, Zeits. f. Physik \u{86}, 120 (1933).
\item[{15.}] B. Rossi, Int. Conf. on Nucl. Phys., London (1934).
\item[{16.}] G. Pfotzer and E. Regener, Nature (London) \u{136}, 718
(1935).     
\item[{17.}] Ref. 2 p. 434. 
\item[{18.}] J.F. Carlson and J.R. Oppenheimer, Phys. Rev. \u{51}, 220
(1937).  
\item[{19.}] B. Rossi, Ric. Scient. \u{1}, vi.559 (1934); \\ P. Auger
and L. Leprince-Ringuet, C.R. Acad. Sci. Paris \u{199}, 785 (1934). 
\item[{20.}] H.J. Bhabha, Proc. Roy. Soc. (London) \u{164A}, 257
(1938). 
\item[{21.}] P.M.S. Blackett, Proc. Roy. Soc. (London) \u{159A}, 1
(1937). 
\item[{22.}] Ref. 20 p. 258.
\item[{23.}] Ref. 20 p. 281.
\item[{24.}] S.H. Neddermeyer and C.D. Anderson, Phys. Rev. \u{51},
884 (1937); \\ J.C. Street and E.C. Stevenson, Phys. Rev. \u{51}, 1004
(1937). 
\item[{25.}] S.H. Neddermeyer and C.D. Anderson, Phys. Rev. \u{54}, 88
(1938).      
\item[{26.}] H. Yukawa, Proc. Phys-Math. Soc. Japan \u{17}, 48
(1935). 
\item[{27.}] H.J. Bhabha, Nature (London) \u{141}, 117 (1938).
\item[{28.}] H.J. Bhabha, Proc. Roy. Soc. (London), \u{166A}, 501
(1938).  See p. 524.   
\item[{29.}] H.J. Bhabha, Proc. Roy. Soc. (London) \u{172A}, 384
(1939).  
\item[{30.}] P.A.M. Dirac, Proc. Roy. Soc. (London) \u{167A}, 148
(1938).  
\item[{31.}] H.J. Bhabha, Proc. Ind. Acad. Sci. \u{11A}, 347, 468
(1940); \\ H.J. Bhabha, Phys. Rev. \u{59}, 100 (1940).
\item[{32.}] H. Snyder, Phys. Rev. \u{53}, 960 (1938); \\ R. Serber,
Phys. Rev. \u{54}, 317 (1938). 
\item[{33.}] H.J. Bhabha and S.K. Chakrabarty,
Proc. Roy. Soc. (London) \u{181A}, 267 (1943) and
Proc. Ind. Acad. Sci. \u{15A}, 464 (1942). 
\item[{34.}]  L. Landau and G. Rumer, Proc. Roy. Soc. (London)
\u{166A}, 213 (1938).
\item[{35.}] K.S.K. Iyengar, Proc. Ind. Acad. Sc. \u{15A}, 195
(1942).  
\item[{36.}] H.J. Bhabha, Proc. Ind. Acad. Sc. \u{19A}, 530 (1944).
\item[{37.}] H.J. Bhabha, S.V. Chandrasekhar Aiya, H.E. Hoteko and
R.C. Saxena, Current Science \u{14}, 98 (1945); Phys. Rev. 68, 147
(1945); and Proc. Nat. Inst. Sci. India \u{12}, 219 (1946).
\item[{38.}] M. Schein, W.P. Jesse and E.O. Wollan, Phys. Rev. \u{47},
207 (1935) and \u{57}, 840 (1940).
\item[{39.}] M.G.K. Menon, in ``Homi Jehangir Bhabha'' 1909-1966, The
Royal Institution of Great Britain, London (1967) p. 17.
\item[{40.}] H.J. Bhabha, Proc. Roy. Soc. (London) \u{202A}, 301
(1950). 
\item[{41.}] H.J. Bhabha and A. Ramakrishnan,
Proc. Ind. Acad. Sc. \u{32A}, 141 (1950). 
\item[{42.}] H.J. Bhabha, Proc. Intl. Conf. on ``Theoretical
Physics'', Kyoto-Tokyo (1953), (Sc. Council of Japan) p. 95 (1954). 
\item[{43.}] H.J. Bhabha and R.R. Daniel, Nature (London) \u{161}, 883
(1948).  
\item[{44.}] B.V. Sreekantan in ``Homi Jahanger Bhabha: Collected
Scientific Papers''.
\item[{45.}] H.J. Bhabha, Proc. Roy. Soc. (London), \u{219A}, 293
(1953). 
\item[{46.}] D.S. Narayan, Nucl. Phys. \u{B34}, 386 (1971). 
\end{enumerate}

\end{document}